\documentstyle[psfig,epsfig]{article}
\newcommand{\mayapr}{\raisebox{-.3ex}{$\enskip\stackrel{>}
{\scriptstyle\sim}\enskip$}}

\begin{document}

\begin{center}
{\LARGE
Towards a Landau--Ginzburg--type Theory \\[5mm] for Granular Fluids}
\\[2cm]
{\large J.~Wakou\footnote[1]{
Institute for Theoretical Physics,
University of Utrecht,
3508 TA Utrecht, The Netherlands},
R.~Brito\footnote[2]{
Departamento  F\'\i sica Aplicada I,
Universidad Complutense,
28040 Madrid, Spain}
and M.H.~Ernst$^1$
}
\\[2.5cm]

\begin{flushleft}
{\it This paper is the contribution of M.H. Ernst to the special issue of 
Journal of Statistical Physics, which contains the Proceedings of 
the 9th International Conference on Discrete Simulation of 
Fluid Dynamics, held on August 21-24, 2000 in Santa Fe, New Mexico}.\\[1cm]
\end{flushleft}

\end{center}

\section*{Abstract}
  
  In this paper we show how, under certain restrictions, 
the hydrodynamic equations for the 
freely evolving granular fluid fit within the framework of 
the time dependent Landau--Ginzburg (LG) models for  critical and 
unstable fluids (e.g. spinodal decomposition). 
  The granular fluid, which is usually modeled as a fluid of inelastic hard 
spheres (IHS), exhibits two instabilities: the spontaneous formation 
of vortices and of high density clusters.
We suppress the clustering instability by imposing constraints 
on the system sizes, in order to illustrate how  LG-equations can be derived
for the order parameter, being the rate of deformation or shear rate tensor, 
which controls the formation of vortex patterns. From 
the shape of the energy functional we obtain the 
stationary patterns in the flow field.
Quantitative predictions of this theory for the stationary 
states agree well with molecular dynamics simulations 
of a fluid of inelastic hard disks.

\section*{Key words} 
Granular fluid; instabilities; pattern formation;
hydrodynamic equations; time dependent Landau-Ginzburg theory.

\section{Introduction}

Granular matter \cite{1} consists of small or large macroscopic
particles. When out of equilibrium, its dynamics is
controlled by dissipative interactions, and distinguished in
quasi-static flows or granular solids on the one hand, and
rapid flows or granular fluids \cite{2} on the other hand.

Typical realizations of granular solids are sand piles,
avalanches, Saturn's rings, grain silos, stress distributions.
Here particles remain essentially in contact, and the dynamics is
controlled by gravity, friction and surface roughness. In this
paper we concentrate on granular fluids. Typical examples are
driven granular flows, such as Poisseuille flow
\cite{losert-gollub}, vibrated beds \cite{swinney,
ScientificAmerican-febr,MORE}, or rapid flows with some form of
continuous energy input \cite{heated-trizac}.
Here the dynamics is controlled by inelastic binary collisions,
separated by ballistic motion of the particles. The forces are
of short range and repulsive, and the system is frequently
modelled as a collection of smooth inelastic hard spheres
(IHS)  \cite{3} of diameter $\sigma$ and mass $m$. Momentum is conserved
during collisions, which makes the system a fluid, but energy is
not conserved. In a collision, on average, a
fraction $\epsilon$ of the relative kinetic energy of the
colliding pair is lost, where $\epsilon$ is referred to as the degree
of inelasticity. In the literature \cite{2,3,4} $\epsilon=1-\alpha^2$,
usually expressed in terms of the coefficient of restitution $\alpha$.
Its detailed definition does not concerns us here.

Here we focus on the idealized limiting case of a freely evolving
rapid flow without energy input and with nearly elastic
collisions, and therefore slowly cooling. This system
shows \cite{4} an interesting instability. When prepared in a
spatially homogeneous equilibrium state, the system does not stay
there, but slowly develops patterns, both in the flow field
(vortices), and in the density field (clusters), the so called
clustering instability.  For the two-dimensional case the analogies 
with spinodal decomposition 
have already been pointed out in the literature~\cite{spinodal}.

The search for the proper macroscopic description of unstable granular fluids     
has been  pursued  by many authors~[1-13,18,20,22-32].
Recently, two new points of view have been presented, namely  by 
Ben-Naim et al.~\cite{ben-naim} and by Soto et al.~\cite{21}.
The first one needs to be discussed in more detail 
because macroscopic equations for granular fluids like 
the Burgers equation appear in that paper, as well as in the present one.
These authors conjecture that the (vector) Burgers equation describes 
the flow velocity ${\bf u}({\bf r},t)$  of a granular fluid, or at least 
of a dilute granular gas --the authors are not very explicit
on this point\cite{WWW-ourcomment}--  on large space and time scales
for {\it arbitrary} values of the inelasticity ($ 0< \epsilon \le 1$) 
in $d-$dimensions, i.e.  
\begin{equation}
\partial_t {\bf u}= -{\bf u}\cdot\nabla{\bf u} +\nu\nabla^2
{\bf u},
 \label{burgers}
\end{equation}
where $\nu$ is the kinematic viscosity, and the solutions of interest
have to satisfy ${\bf \nabla} \times {\bf u}= {\bf 0}$. 
An interesting aside is that the rotation-free solution of 
the vector Burgers equation can be expressed as the gradient 
of a scalar field, ${\bf u} =\nabla h({\bf r},t)$, 
where $h({\bf r},t)$ satisfies the equation,
\begin{equation}
\partial_t h = \nu \nabla^2 h + \textstyle{\frac{1}{ 2}}|\nabla h|^2.
 \label{KPZ}
\end{equation}
This equation is the famous KPZ-equation, named after Khadar,
Parisi and Zhang \cite{ref to KPZ}. It  describes the growth dynamics
of solid surfaces, where $h$ is the height function.

It is well known that the Burgers 
equation in the inviscid limit ($\nu \to 0$) describes the "sticky dust"
or adhesion model of {\it perfectly inelastic point} particles with
$\epsilon =1$ \cite{17,18,19}.
The conjectured validity of the Burgers equation  for the inelastic hard sphere 
fluid implies a universality hypothesis, i.e. the large 
$({\bf r},t)-$behavior of a granular gas is in the universality class of  sticky 
dust, independent of its inelasticity $\epsilon$.
Several implications of this  conjecture in dimensions $d\ge 2$
have been criticized in Ref~\cite{trizac-barrat,WWW-ourcomment}. 
For the one-dimensional case the authors present rather convincing evidence 
based on molecular dynamics (MD) simulations using $N=10^6$ particles, and 
on scaling arguments. 
In addition, Boldyrev~\cite{boldyrev} has studied the one-dimensional 
randomly driven Burgers equation with a pressure term
$-(\nabla p)/\rho$ with $p\propto n^a$ included.
He has shown that properties, like the structure factor or pair correlation 
function, are not affected by the pressure term
if $a<2$. On account of this argument, 
the $\nabla p$-term in the one-dimensional Navier-Stokes equation
can be neglected on the largest $(r,t)-$ scales
for a dilute gas in the inviscid limit, making the Burgers equation 
the appropriate macroscopic equation. Note that in the freely cooling 
granular fluid the viscosity decreases as $\sqrt{T}$ as $T\to 0$.
In higher dimensions the systems used in MD simulations 
($N=5 \times 10^4$ in two dimensions~\cite{WWW-ourcomment} and $N=10^6$ 
in three dimensions~\cite{chen-physlett}) are too small to draw any conclusions 
on large 
scale behavior. Here analytic or scaling arguments to support the
conjecture are lacking.

Although equations of the form (\ref{burgers}) will frequently appear 
in the present paper, we emphasize that Ref.~\cite{ben-naim} refers only to 
the largest possible scales (where the thermodynamic limit has been taken). 
In this paper we explicitly restrict ourselves to small systems 
in order to suppress the clustering instability. 
This has been done  to simplify the problem.
Moreover, Ref.~\cite{ben-naim} only refers to 
dilute granular gases, whereas this paper covers also 
liquid densities.

A second new development, which is somewhat similar to ours, is
given by Soto et al.~\cite{21}. These authors also study 
granular fluids, contained in systems that are sufficiently small,
such that the clustering instability is suppressed.
In these small systems the growth of vortices is very slow.
Consequently, the remaining hydrodynamic modes are enslaved
by the slowly growing unstable vorticity modes
and the amplitude of these vorticity modes remains very small.
Under this condition, they obtained the amplitude equations
for the slowest vorticity modes, which can be derived from a
potential function.
 
The question of interest in the present paper is: can the models
 for granular fluids be fitted into the
generic classification of Landau--Ginzburg--type models, as given
by Hohenberg and Halperin~\cite{5}, to describe critical dynamics
and hydrodynamic instabilities? The goal of this article is to
illustrate how, under certain restrictions, the standard nonlinear
hydrodynamic equations for the IHS fluid \cite{2,3} can be cast
into a Landau--Ginzburg--type equation of motion for the order
parameter, which can be derived from an energy functional and,
more specifically, to point out which terms in the original
hydrodynamic equations are responsible for the quartic terms in
the Landau-Ginzburg energy functional.   

The plan of the paper is as follows. In Sec.2, we start with 
the hydrodynamic equations. The decay of the total energy
at short times and the results of a linear stability analysis
are briefly reviewed. In Sec.3, we introduce an assumption
of incompressible flows under  certain restrictions on
system size or time regime.
Then, under these assumptions, the hydrodynamic equations
are reduced to a closed equation for a scaled flow field.
It is shown in Sec.4 that this equation for a scaled flow field
can be cast into the form of a time--dependent 
Landau--Ginzburg equation for
an appropriate order parameter. The shape of the energy functional
is discussed and possible stationary solutions are presented.
Finally, in Sec.5, we make a quantitative comparison of the theoretical 
predictions at large times  with molecular dynamics simulations of inelastic 
hard disks. We end with some conclusions in Sec.6.

\section{Dynamic Equations and Instabilities}

The starting point are the hydrodynamic equations. They are not
only needed here to recapitulate our present theoretical
understanding of this system, but also to formulate the new
extensions to be discussed in this paper.

 The macroscopic time evolution of the IHS fluid on large
spatial and temporal scales \cite{2,4} can be described by the
nonlinear hydrodynamic equations for the local density $n({\bf
r},t)$, the local flow field ${\bf u} ({\bf r},t)$ and the local
temperature $T({\bf r},t)$, supplemented with a sink term $\Gamma$
accounting for the energy loss through inelastic collisions, i.e.
\begin{eqnarray}
D_t n       &=& -n\nabla\cdot{\bf u}, \nonumber \\
D_t {\bf u} &=& -\frac{1}{mn} \nabla p + 2 \nu \nabla\cdot
{\mbox{\sf D}}, \nonumber \\
D_t T       &=& -\frac{2p}{dn} \nabla\cdot{\bf u}
+ b_T \nabla^2T
+2 b_\perp {\mbox{\sf D}}:{\mbox{\sf D}} -\Gamma .
\label{NS}
\end{eqnarray}
 In this paper the inelasticity is always assumed to be small.
For later convenience the macroscopic equations are given for a
$d$-dimensional systems. Here
$D_t=\partial_t+{\bf u} \cdot\nabla$ is the time derivative in a
comoving frame. The local energy density of the IHS fluid is
$e=\frac{1}{2}mnu^2 +\frac{d}{2}nT$, and $p$ is the pressure. The
shear rate $D_{\alpha\beta}$  is the symmetrized dyadic, $\{
\nabla_\alpha u_\beta \}$, which is also made traceless. The
coefficient $b_T=2\kappa/dn$ is proportional to the heat
conductivity $\kappa$, and $b_\perp=2m\nu/d$ to the shear
viscosity $\nu$. For simplicity of presentation the bulk viscosity
has been set equal to zero, and the transport coefficients $\nu$
and $\kappa$, which depend on the local density and temperature,
are taken at some fixed reference values, $\bar{n}(t)$ and
$\bar{T}(t)$, to be specified later. The four terms in the energy
balance equation account for work done by the pressure, for heat
conduction, for nonlinear viscous heating and collisional dissipation.
Gradients in the flow velocity considerably slow down the
collisional cooling process through viscous heating.

On the basis of kinetic theory one can derive that the rate of
collisional energy loss, $\Gamma = 2 \gamma_0\omega T$, is
proportional to the collision frequency $\omega$ multiplied by the
fraction of energy $\epsilon T$ lost per collision \cite{8,jago-boston},
where $  \gamma_0   =\epsilon /2d = (1-\alpha^2)/2d $. In general,
the collision frequency  $\omega(T)$ is proportional to the root
mean square velocity $v_0=\sqrt{2T/m}$, and its explicit form for
hard sphere fluids can be found in Refs.~\cite{11,12}. When the
system is prepared initially in a homogeneous equilibrium state,
it evolves at short times in a spatially homogeneous cooling state
(HCS) with a time dependent  temperature. Combination of
Eqs.~(\ref{NS}) and the expression for $\Gamma$ given above,
yields $\partial_t T\sim
-\gamma_0 T^{3/2}$. This leads to Haff's homogeneous cooling law~\cite{10} 
for the mean energy per particle in the IHS fluid,
\begin{equation}
E(t)=\frac{d}{2} T(t)  = \frac{E_0}{(1+\gamma_0\omega_0 t)^2}
     =E_0 \exp ( -2 \gamma_0 \tau ),
\label{haffslaw}
\end{equation}
which is needed for later comparison. Here $E_0=(d/2)T_0$ is the
initial energy, and $t_0=1/\omega_0$ with $\omega_0=\omega(T_0)$
is the mean free time in the initial state. In general the number
of collisions per particle $\tau(t)$ in a time $t$  is defined
through $ d \tau = \omega (T(t)) d t$. In the  HCS integration of this
relation  yields  $ \exp( \gamma_0 \tau ) =
(1+\gamma_0\omega_0 t)$.

However, this state is unstable against spatial fluctuations in
density $n({\bf r},t)$, temperature $T({\bf r},t)$ and flow velocity ${\bf
u}({\bf r},t)$. The present theoretical understanding of these instabilities
\cite{deltour,16,esipov,jago-boston,12}  is
essentially based on a linear stability analysis of the
hydrodynamic fluctuations in the density, $\delta n = n -{\bar
n}$, temperature, $\delta {T}= T -{\bar T} $, and flow velocity
${\bf u}$. This is done by using the rescaled Fourier modes
$\delta n_{\bf k}$, $\delta {\tilde{T}_{\bf k}}= \delta T_{\bf k}/{\bar T}$ and
$\delta {\tilde{ {\bf u}}_{\bf k}} \sim {\bf u_k}/ \sqrt{{\bar T}}$,
where an overline denotes a spatial average, ${\bar a} = (1/V)
\int_V d {\bf r}\, a({\bf r})$, and ${\bar T}$ is the global granular
temperature.
Fourier transforms are defined as 
$f_{\bf k}=
\int_V d {\bf r}\, e^{-i {\bf k}\cdot {\bf r}} f({\bf r})$.
 The rescaled eigenmodes are described by $\delta
a_{\bf k}(\tau) = \exp(z_\lambda \tau) \delta a_{\bf k}(0)$.  The exponential
growth rates of unstable ($z_\lambda (k)>0 $) and stable
($z_\lambda (k) < 0 $) modes are shown in Fig.~1 as a function of
the wave number $k$.

 This figure shows that the
transverse flow field ${\bf u}_{\perp{\bf k}}$ or shear mode
($\lambda=\perp$) with a wave number $k<k^*_\perp$ is unstable, and
develops vortices. On the other hand density fluctuations couple
weakly, in order ${\cal O}(k)$, to the heat modes ($\lambda=H$),
and $z_H(k)$ in Fig.1 shows that these fluctuations are unstable
in the range  $k<k^*_H$, and linearly stable in the range $k>k^*_H$ , i.e.
remain at thermal noise level.  The stability thresholds
$k^*_\perp$ and $k^*_H$ are defined as the the root of
$z_\lambda(k)=0$ for $\lambda=\{\perp,H\}$, and are marked as
black dots in the figure. The figure also shows that the growth
rate $z_\bot (k)$ for the vorticity mode is much larger than
the growth rate for the heat mode $z_H (k)$, which couples to the
density fluctuations. This explains why vortices appear long
before the density clusters start to appear.

An intuitive explanation for the  appearance and growth of large
scale vortices is that a binary collision destroys a fraction of
the kinetic energy of relative motion of the colliding pair. The
cumulative effect of many successive collisions is that they make
the particles move  locally in a more parallel and coherent
fashion. This creates locally patches of vorticity. These patches
grow in size by selective suppression (stronger damping) of short
wavelength modes~\cite{EPL-brito}.

\begin{figure}
\begin{center}
    \epsfig{file=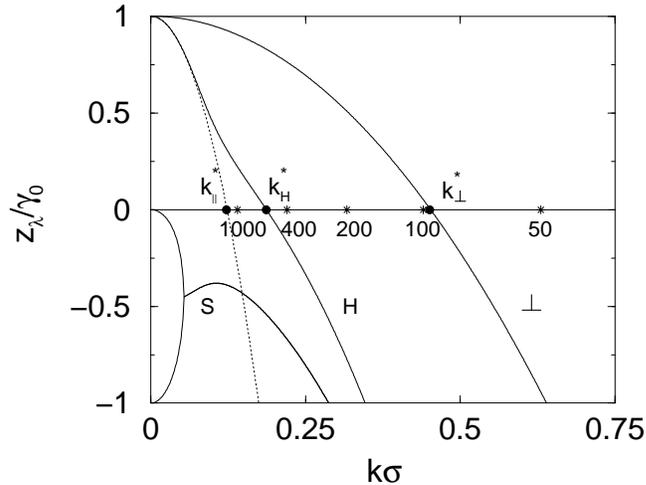,angle=270,width=8.5cm}
\end{center}
\caption{\small Dispersion relations $z_\lambda(k)$ (from right to
left) for the shear mode $(\lambda = \perp)$, heat mode $(\lambda
=H)$ and the sound modes in the IHS fluid for an area fraction of
$\phi=0.4$ and a restitution coefficient $\alpha=0.85$. The stars
mark the location of the  minimum wave vector allowed in the
system, $k_0=2\pi/L$, for  the number of particles indicated, and
the black dots mark the location of the threshold values $k^*_a
\sim 1/\xi_a$ with corresponding correlation lengths $\xi_a$ where
$a= (\parallel,H,\perp)$.}
\end{figure}

An intuitive explanation of the appearance of high density
clusters goes as follows \cite{4}.
Suppose one prepares the system
initially in a spatially homogeneous equilibrium state, and there
occurs locally a spontaneous negative pressure fluctuation
(`depression'). The resulting particle flow from the surroundings
tries to compensate for the local depression. This creates an
excess local density, which increases the collision frequency, and
in turn decreases the temperature. This creates again a
depression, and the process keeps repeating itself, thus creating
cold dense clusters, surrounded by a hot dilute gas. Moreover,
these arguments also suggest that the pressure fluctuations and
the corresponding gradients remain substantially smaller than
those in density and temperature. In fact it would be very
interesting if one could test this suggestion by measuring the
pressure {\it locally}, i.e.~in cells of the typical size of a
density cluster, through molecular dynamics simulations, or Direct
Monte Carlo Simulations (DSMC) of the Boltzmann or Enskog
equation, or by the Lattice Boltzmann method.

The shape of  the dispersion relations for the growth rates also
explains the suppression of instabilities through a reduction of
the system size \cite{deltour,16,esipov,jago-boston}.
 Let $k_0 (N) = 2\pi/L$ be the smallest wave number,
allowed in a system of linear extent $L$ at fixed density and 
with periodic boundary
conditions. In systems with $k^*_H<k_0(N)<k^*_\perp$, there are
growing vorticity modes, but all density fluctuations are stable
according to a linear stability analysis. In systems with
$k_0(N)<k^*_H$ the fluctuations in the density and in the flow
field are unstable~\cite{deltour}. 
However, systems with $k_0(N)>k_\perp^*$  do not
show any instability.
The smallest allowed wave numbers
$k_0(N)$ for different numbers of particles, $N=50,\, 100,\,
200,\dots,\,1000$ at fixed packing fraction $\phi$, are  shown as
stars in Fig.1. This information on small systems will be used in
later sections.

We finally remark that several authors  have also studied
nonlinear terms in the macroscopic equations for granular fluids,
such as the viscous {\it heating} term
\cite{4,brey-mariajose,21,wakou}, and the nonlinear convective term
\cite{ben-naim}. The inclusion of the combined effects of viscous
heating and collisional cooling is essentially the new mechanism
driving the dynamics of dissipative granular fluids in the time
regime, directly following the homogeneous cooling state.

In an unpublished report \cite{wakou} one of us has developed a
systematic expansion method for solving the nonlinear
hydrodynamic equations for granular fluids. The method is based on
the separation of time scales of the relatively slow process of
vorticity diffusion and the relatively fast process of heat
conduction for all $k$-dependent excitations, allowed in the
system ($k > k_0(N)$). This picture is consistent with the
relative sizes of the growth rates $z_{\perp}(k) -\gamma_0$ and
$z_H(k) -\gamma_0$, as shown in Fig. 1.

The separation of the time scales permits a systematic
approximation scheme, based on the Chapman-Enskog method or
multi-time scale expansion method~\cite{haken,chapman-cowling}. 
In zeroth order this method yields a closed
equation  of motion for the rescaled flow field ${\bf \tilde{u}}$,
as well as an equation for the global granular temperature ${\bar
T}(t)$.  In first approximation (see Ref.~\cite{wakou}) this method
yields equations for the density and temperature fluctuations,
$\delta n$ and $\delta T$, and a higher order correction to the
Navier-Stokes equation.

In the present paper we will not explain the rather technical
calculations based on the multi-time scale method, but we will try
to elucidate the essential physics by deriving the zeroth order
results in a more intuitive fashion. The systematic derivation
will be published elsewhere.

\section{Incompressible flows}

As discussed in the previous section, instabilities and pattern
formation occur in two different local fields, ${\bf  u}$ and $n$,
and on two different time scales, namely first patches of
vorticity appear, and only much later density clusters appear. As
explained above, the appearance of density clusters can even be
further delayed, or all together suppressed  by decreasing the
system size \cite{deltour,16,esipov,jago-boston}. These observations 
suggest to 
analyze first the simplest case where fluctuations in density and
temperature remain small. This would happen in the time regime
following the unstable homogeneous cooling state, and possibly in
the full time regime for sufficiently small systems, as suggested
by the linear stability analysis. Of course the stability of
solutions on the largest time scales is determined by the full
nonlinear equations. The conditions for small $n-$ and
$T-$fluctuations might be realized in {\it incompressible} flows,
where ${\bf \nabla \cdot u} = 0$. Then the density remains
constant in the comoving frame, and the temperature balance
equation greatly simplifies. As is well known from standard fluid
dynamics and from the theory of turbulence \cite{13,14}, flows of elastic
fluids are quite {\em incompressible}. This implies,
\begin{equation}
\nabla\cdot{\bf u}=0 \quad {\rm or} \quad
{ u}_{\parallel{\bf k}} \equiv \hat{\bf k}\cdot
{\bf u}_{\bf k}={ 0},
\label{upar0}
\end{equation}
where ${ u}_{\parallel{\bf k}}$ is the longitudinal Fourier mode.
Moreover, MD simulations and the theory of hydrodynamic
fluctuations in granular flows \cite{11,12} show that the
incompressibility assumption, ${u}_{\parallel{\bf k}}=0$,
remains valid down to very small wave numbers, satisfying
$k\xi_\parallel \mayapr 1$ , and ultimately breaks down at the
largest wavelengths, where $\xi_\parallel \sim 1/\gamma_0$ is the
largest intrinsic dynamic correlation length in IHS fluid. It
satisfies the inequality, $\xi_\parallel\gg \xi_\perp \equiv
(\nu/\omega\gamma_0)^{1/2}$ for nearly elastic systems. Both
correlation lengths are indicated in Fig.1 and defined more
explicitly in \cite{12}.

Therefore, as a zeroth approximation to our nonlinear theory, we
make the incompressibility assumption,
$\nabla\cdot{\bf u}=0$, following Refs.~\cite{11,12}, and the
 equations of motion become,
\begin{eqnarray}
D_t n       &=& 0, \nonumber \\
D_t {\bf u} &=& -\frac{1}{mn} {\bf \nabla} p +\nu \nabla^2 {\bf
u}, \nonumber \\
D_t T       &=&  b_T \nabla^2T
+ b_\perp \left[\nabla{\bf u} + (\nabla{\bf
u})^\dag\right]:\nabla{\bf u}  - 2\gamma_0\omega T.
\label{NSI}
\end{eqnarray}
Consequently $n({\bf r},t)$ is constant in a comoving frame, and
the set of nonlinear equations (\ref{NSI}) can not describe the
growth of inhomogeneities in the density field, but supposedly describe
the time evolution of the system as long as the density
fluctuations are small.

As a consequence of the incompressibility assumption the local
density stays  constant in time, and neither depressions, nor hot
and cold regions can develop. Therefore, the pressure gradient in
(\ref{NSI}) remains negligibly small as well, and the
Navier-Stokes equation becomes,
\begin{equation}
\partial_t {\bf u}= -{\bf u}\cdot\nabla{\bf u} +\nu\nabla^2
{\bf u},
 \label{NS2}
\end{equation}
where the flow velocity ${\bf u} = {\bf u_\bot}$ is purely
rotational with $\nabla \cdot {\bf u}=0$. We also note that 
the divergence of Eq.(\ref{NS2}), in combination with 
$\nabla \cdot {\bf u}=0$, reduces to 
$(\nabla_\alpha u_\beta)(\nabla_\beta u_\alpha) =0$ at all times.
As already discussed in the introduction, the Burgers equation
for a rotation-free flow field has been conjectured~\cite{ben-naim}
as the large scale macroscopic equation for a d-dimensional 
dilute granular gas of arbitrary inelasticity $\epsilon$. 

Next we consider the temperature balance equation, which
involves two processes: the diffusion process of
heat conduction, where Fourier modes $T_{\bf k}$ decay with a rate
$b_T k^2$, and the global process
of collisional cooling and nonlinear viscous heating,
describing the decay of $T_{\bf k}$ for $k \to 0$,  or
equivalently of $\bar{T}(t)
\equiv (1/V) \int_V d {\bf r}\, T({\bf r},t)$, referred to
as global temperature.

To split off the behavior of $\bar{T}(t)$ from Eq.(\ref{NSI})
we take the spatial average of the $T$--equation, yielding
for the global temperature,
\begin{equation}
\partial_t \bar T=  b_\perp \overline{|\nabla{\bf u}|^2}
-2\gamma_0\omega \bar T,
\label{Tbar}
\end{equation}
where overlines denote spatial averages.
We have introduced the notation $|{\mbox{\sf A}}|^2=
\sum_{\alpha\beta}|A_{\alpha\beta}|^2$, for a second rank tensor
${\mbox{\sf A}}$.
The transport coefficient $b_\perp$ and collision frequency
$\omega$ are functions of the spatially average density
$\bar {n}= N/V$ and temperature ${\bar T}(t)$ (see 'reference values'
below Eq.~(\ref{NS})).
This is allowed as long as the local fluctuations $\delta n = n -\bar n$
and $\delta T= T -\bar T$ are not getting too large through
nonlinear effects.

The new evolution equations~(\ref{NS2}) and (\ref{Tbar}) contain the
time dependent coefficients $ b_\bot$, $ \omega$ and $\nu$, which
are proportional to $ \bar{v}_0(t)\equiv ( 2\bar T(t)/m)^{1/2}$.
Therefore, it is convenient to introduce the scaled field $\tilde{\bf u}
= {\bf u}/{\bar{v}}_0$, and the scaled time $\tau$, defined as
$d\tau=\omega( \bar T) d t$. The final macroscopic evolution
equations then become,
\begin{eqnarray}
\partial_\tau\ln\bar T &=& \frac{4}{d} {\cal D}_\perp
\overline{|\nabla\tilde{\bf u}|^2} -2\gamma_0, \nonumber \\
\partial_\tau \tilde{\bf u} + l_0\,\tilde{\bf u} \cdot \nabla \tilde{\bf u}
&=& {\cal D}_\perp
\nabla^2 \tilde{\bf u} -\frac{1}{2} (\partial_\tau\ln \bar T ) \tilde
{\bf u} \nonumber \\
&=& \gamma_0 \tilde{\bf u} + {\cal D}_\perp \nabla^2 \tilde{\bf u}
-\frac{2}{d} {\cal D}_\perp \overline{|\nabla\tilde{\bf u}|^2}
\tilde{\bf u},
\label{reduced}
\end{eqnarray}
where $l_0=\bar{v}_0/\omega$ is the (time-independent) mean free path.
The last equation is a closed equation for $\tilde{\bf u}$, and the
physically consistent solutions need to obey the relation
$\nabla\cdot{\bf u}=0$. The global temperature is slaved by the flow field.
The rescaled vorticity diffusion coefficient,
defined as ${\cal D}_\perp = \nu/\omega$, is independent of time. The first term
on the right hand side of the equation for $\tilde{\bf u}$
accounts for the instability, the second for the vorticity diffusion and the
last
one for the saturation effects, caused by the nonlinear viscous
heating. It slows down the growth of unstable $\bf k$-modes,
and may ultimately lead to a steady state for $\tilde{\bf u}$.
For large times the above equations (\ref{NS2}) and (\ref{Tbar})
are no longer valid, and nonlinear effects will induce density
inhomogeneities, even in small systems with
$k^*_H<k_0(N)<k^*_\perp$.

\section{Spontaneous symmetry breaking}

In this section we will drop the nonlinear convective term
$\tilde{\bf u} \cdot {\bf \nabla} \tilde{\bf u}$, but at
the end of our analysis we admit out of all possible solutions
only those that satisfy Eq. (\ref{reduced}) with the
convective term included.

The final equation for the rescaled ${\bf u}-$field has the form of the
Landau-Ginzburg equation of motion for a {\it non-conserved} order parameter.
This can be made more explicit by  introducing the order parameter
 ${\mbox{\sf S}} =\nabla\tilde{\bf u}$, and applying
$\nabla$ to the ${\bf u}$-equation in (\ref{reduced})
with the result,
\begin{eqnarray}
\partial_\tau {\mbox{\sf S}} &=& \left(\gamma_0+
{\cal D}_\perp\nabla^2 \right)
{\mbox{\sf S}} -\frac{2}{d} {\cal D}_\perp
\overline{|{\mbox{\sf S}}|^2} {\mbox{\sf S}} \nonumber \\
&=& -V \delta {\cal H} [{\mbox{\sf S}}] /\delta {\mbox{\sf S}}^\dag,
\label{landau}
\end{eqnarray}
where the last line contains the functional derivative of the
energy functional ${\cal H} [{\mbox{\sf S}}]$, defined as
\begin{equation}
{\cal H} [{\mbox{\sf S}}] = -\frac{1}{2} \gamma_0
\overline{|{\mbox{\sf S}}|^2} + \frac{1}{2} {\cal D}_\perp
\overline{|\nabla{\mbox{\sf S}}|^2} + \frac{1}{2d}
{\cal D}_\perp \left(\overline{|{\mbox{\sf S}}|^2}\right)^2.
\label{landauf}
\end{equation}
In our further considerations it is more convenient to use
Fourier modes. Moreover
${\bf u}_{\bf k}={\bf u}_{\perp\bf k}$, because of the assumption
of incompressibility, and ${\mbox{\sf S}}_{\bf k}\equiv
{\bf k}\tilde{\bf u}_{\perp\bf k}$. Then we obtain the equation of
motion,
\begin{eqnarray}
\partial_\tau {\mbox{\sf S}}_{\bf k} &=&
-V^2 \delta {\cal H}[{\mbox{\sf S}}]/\delta {\mbox{\sf S}}_{\bf k}^\dag
\nonumber \\
&=& \left\{\gamma_0-{\cal D}_\perp k^2-\frac{2 {\cal D}_\perp} { d V^2}\sum_{\bf
q} |{\mbox{\sf S}}_{\bf q}|^2\right\}
{\mbox{\sf S}}_{\bf k}
\label{Sk}
\end{eqnarray}
with an energy functional,
\begin{equation}
{\cal H} [{\mbox{\sf S}}] = \frac{1}{2V^2} \sum_{\bf k}
(-\gamma_0  + {\cal D}_\perp k^2) |{\mbox{\sf S}}_{\bf
k}|^2 + \frac{{\cal D}_\perp}{2d} \left(\frac{1}{V^2}\sum_{\bf k} |{\mbox{\sf
S}}_{\bf k}|^2 \right)^2,
\label{landaur}
\end{equation}
where the wave number ${\bf k}={\bf 0}$ does not contribute.

The energy in (\ref{landauf}) and (\ref{landaur})
 resembles a Landau free energy form for a {\em tensorial order
parameter} ${\mbox{\sf S}} =\nabla\tilde{\bf u}$ with
tr\,${\mbox{\sf S}}=\nabla\cdot\tilde{\bf u}=0$, which is in fact
the shear rate or rate of deformation tensor. It has a quartic
term $ S^4$, and a quadratic term  $S^2$ with a coefficient that
vanishes at a critical threshold $k^*_\bot = (\gamma_0/{\cal
D}_\bot)^{1/2}$. It differs from the standard Landau free energy in that
the quartic term contains summations over two totally independent wave numbers.

These results are very interesting. If the energy functional has a
minimum, then there is a fixed point solution, ${\mbox{\sf S}}_{\bf
k}(\infty)$, that is approached for large times. They are found by setting the
right hand side of (\ref{Sk}) equal to zero, i.e.,

\begin{equation}
\left\{\gamma_0-{\cal D}_\perp k^2-
\frac{2 {\cal D}_\perp} { d V^2}
 \sum_{\bf q} |{\mbox{\sf S}}_{\bf q}|^2\right\}
{\mbox{\sf S}}_{\bf k}=0.
\end{equation}

\begin{figure}
\begin{center}
          \epsfig{file=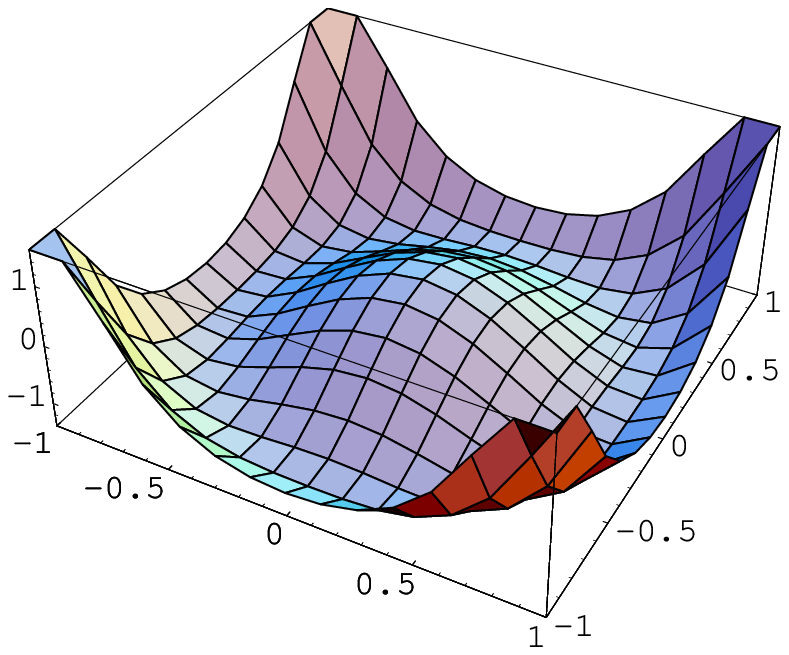,width=8cm}
\end{center}
\caption{\small Landscape plot of the energy ${\cal H}
[{\mbox{\sf S}}]$ in the ${\mbox{\sf S}}_{{\bf k}_0}-$subspace.}
\end{figure}

We will show that depending on the  parameter, $k_0=2\pi/L$,
being above or below the threshold value $k^*_\perp
\equiv \sqrt{\gamma_0/D_{\perp}}$, the
fixed point value of the
order parameter, $\{{\mbox{\sf S}}_{\bf k}(\infty)\}$, is
 {\em vanishing} or {\em non-vanishing}.
 A stable fixed point with some non-vanishing Fourier
components indicates that the system approaches a non-equilibrium
steady state with a stationary pattern in the flow field, and
spontaneous symmetry breaking has occurred.

Consider the right hand side of (\ref{Sk}), and observe that the
expression between curly brackets is necessarily {\em negative}
for $k>k^*_\perp $, and the Fourier mode ${\mbox{\sf S}}_{\bf k}$
decays to zero. If the smallest possible wave number satisfies,
$k_0 > k^*_\perp$, {\em all} ${\mbox{\sf S}}_{\bf k}$ decay to
zero, and there is no spontaneous symmetry breaking. The system
remains spatially homogeneous. However, if $k_0<k^*_\perp$, then
there exists the possibility that the expression inside brackets
in Eq.(\ref{Sk}) vanishes for a non-vanishing value of ${\mbox{\sf
S}}_{\bf k}(\infty)$, i.e.~there is an extremum determined by the
condition,
\begin{equation}
\frac{1}{V^2}\sum_{\bf q} |{\mbox{\sf S}}_{\bf q}|^2 =\frac{d}{2}
(\gamma_0  - {\cal D}_\perp k^2) /{\cal D}_\perp.
\label{velocityamp}
\end{equation}
If the extremum is a saddle point then there are directions of
exponentially growing solutions. However, if $k\le k^*_\perp$, then
the right hand side of~(\ref{velocityamp}) may become zero and
even positive. There is the possibility of stationary and of
exponentially growing solutions. The fixed point $\{{\mbox{\sf
S}}_{\bf k}(\infty)=0 \mbox{ for any ${\bf k}$ } \}$
is a saddle point. One can also show \cite{wakou} that all fixed point
solutions with non-vanishing ${\mbox{\sf S}}_{\bf k}$ for any
$|{\bf k}|\neq k_0$ are saddle points, and that the only
non-vanishing solution
 $\{{\mbox{\sf S}}_{\bf k}\ne 0 \mbox{ if }|{\bf k}|=k_0,
{\mbox{\sf S}}_{\bf k}=0\mbox{ otherwise}\}$ is a stable fixed
point with an infinitely degenerate minimum, given symbolically by
the Mexican hat shape as illustrated in Fig. 3. The condition
(\ref{velocityamp}) for ${\bf u_{\perp {\bf k}}}$ with $|{\bf
k}|=k_0=2\pi/L$ in 
$d$-dimensions is then
\begin{equation}
\frac{1}{V^2}\sum_{\alpha=1}^{d}
\left|\tilde{\bf u}_{{\bf k}_{0\alpha}}\right|^2
=
\frac{d}{4}(\gamma_0-{\cal D}_\perp k_0^2)/{\cal D}_\perp
k_0^2 ,
\label{velstationary}
\end{equation}
where ${\bf k}_{0\alpha}=k_0\hat{\bf e}_{\alpha}$ and 
$\hat{\bf e}_{\alpha}$ is a unit vector in the direction $\alpha$
($\alpha=\{1,2,\cdots d\}$).
As the solutions of these equations have to satisfy Eq.~(\ref{upar0}),
the Fourier components can be written as 
\begin{eqnarray}
\tilde{\bf u}_{{\bf k}_{0\alpha}}
=
\frac{1}{2} V \sum_{\beta(\ne\alpha)}
\hat{\bf e}_{\beta} A_{\alpha \beta} \,e^{i\,\theta_{\alpha \beta}}
,
\label{k0modes}
\end{eqnarray}
where the phases $\theta_{\alpha \beta}$ and amplitudes $A_{\alpha \beta}$
are $2d(d-1)$ real numbers. On account of (\ref{velstationary}) the 
amplitudes satisfy the relations
\begin{eqnarray}
A_{0}^2
\equiv
\mathop{\sum\sum}_{1\le\alpha\ne\beta\le d}
\,A_{\alpha \beta}^2
=
d(\gamma_0-{\cal D}_\perp k_0^2)/{\cal D}_\perp
k_0^2 
.
\label{amplitudes}
\end{eqnarray}
The stable stationary solutions in real space are then
\begin{eqnarray}
\tilde{\bf u}_0({\bf r})
=
\mathop{\sum\sum}_{\alpha\ne\beta} 
A_{\alpha\beta}\,\hat{\bf e}_{\beta}
\cos\left(k_0 r_{\alpha}+\theta_{\alpha\beta}\right).
\label{statsolu}
\end{eqnarray}
Out of this set of solutions we select
those that satisfy the full nonlinear Eq.~(\ref{reduced})
with the convective term included, i.e. we determine
the $d(d-1)$ amplitudes $A_{\alpha\beta}$ such that 
the relation, $\tilde{\bf u}_0\cdot\nabla\tilde{\bf u}_0={\bf 0}$,
is satisfied. This yields $d$ conditions.
For the two-dimensional case only {\it two} fixed
point solutions remain, instead of infinitely many
degenerate minima, i.e.
\begin{eqnarray}
\tilde{\bf u}_0(x)  &=&  \hat{\bf e}_y A_0 \cos(k_0 x+\theta_x)
\nonumber \\
\tilde{\bf u}_0(y)  &=&  \hat{\bf e}_x A_0 \cos(k_0 y+\theta_y).
\label{us}
\end{eqnarray}
The symmetry of the
steady state is spontaneously broken, and
  spontaneous fluctuations in the initial state determine
which of these two minima will be reached.

In summary, the equation, describing the growth dynamics of vortices 
in granular fluids, {\it without} the convective term is described by 
a time dependent Landau-Ginzburg equation for a non-conserved 
order parameter, ${\mbox{\sf S}} =\nabla\tilde{\bf u}$,
derived from an energy functional with a {\it continuous} set of degenerate
 minima, having the shape of a Mexican hat.  The last observation
may suggest that unstable granular fluids 
have some resemblance to Model H~\cite{5}
which has a similar energy surface in the neighborhood of
its stable fix points. However, this is not the case. 
Addition of the nonlinear convective term to Eq.(\ref{landau}) 
selects out of this infinite set of minima only
subsets of admissible solutions. In two dimensions only {\it two} 
distinct minima survive. Therefore, the unstable two-dimensional IHS 
fluid has a greater resemblance to spinodal decomposition for 
a non-conserved order parameter, which belongs to the universality class of 
Model A~\cite{5}.

It should be noted that the complete solution of Eq.~(\ref{reduced})
with the convective term included may have a larger set of physically
acceptable solutions. However, we have not been able to find any.

\section{ MD-Simulations}
In this section we present results obtained from the zeroth 
approximation and compare the theoretical predictions with the results
of computer simulations of small systems in two dimensions. 
At the end of this section, some results of the first 
order approximation in Ref.~\cite{wakou} are presented without
derivation and compared with computer simulations.

\begin{figure}[h]
\begin{center}
\begin{minipage}{12cm}
\begin{minipage}{7cm}\epsfig{file=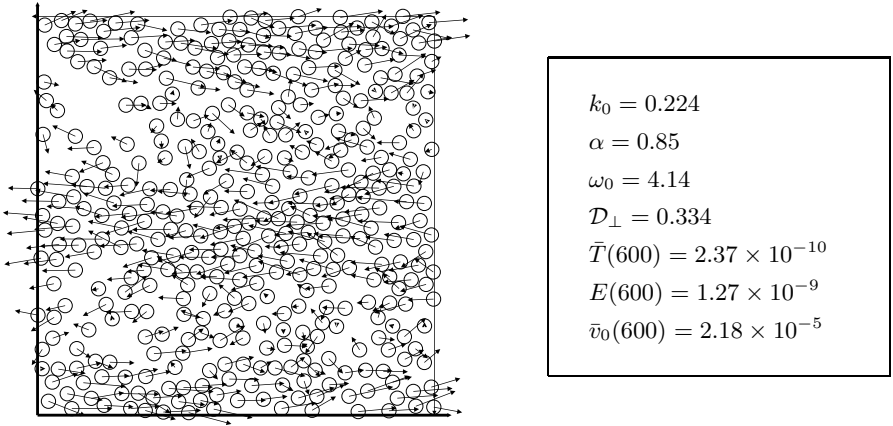,width=6cm} \end{minipage}
{\small
\begin{tabular}{|c|}
\hline \ \ \
\begin{minipage}{3.8cm}\ \\[2mm]
$ k_0=0.224$                             \\
$ \alpha=0.85 $                          \\
$ \omega_0=4.14$                         \\
$ {\cal D}_\bot =0.334 $                 \\
${\bar T}(600) =2.37\times 10^{-10}$     \\
$E(600)= 1.27\times 10^{-9}$             \\
${\bar v}_0 (600) = 2.18\times 10^{-5} $ \\ \
\end{minipage} \\
\hline
\end{tabular}
}
\end{minipage}
   \end{center}
\caption{\small Left: Configuration with $N=400$ at $\tau=600$.
A shear flow
is observed in agreement with the solution of the nonlinear
equations given in Eq.~(\protect{\ref{us}}).
Right: Numerical data for this and subsequent plots.
Last three entries are at $\tau=600$. Units are chosen such that
initial temperature $T_0=1$, mass $m=1$, and sphere diameter $\sigma=1$.}
\label{config}
\end{figure}

\begin{figure}[h]
\begin{center}
      \epsfig{file=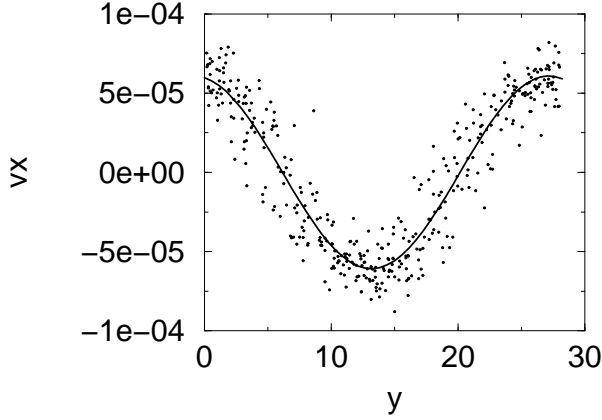,width=8cm} \hspace*{1.4cm}
\vspace*{-0.5cm}
\end{center}
\caption{\small Profile of shear flow in Fig.\ref{config}. 
Dots represent
the instantaneous velocities of the
particles; the solid line is a fit to the sinusoidal
function, described in Eq.~(\protect{\ref{us}}),
with amplitude $A_{\mbox{sim}}/\bar{v}_0\simeq 2.78$
and theoretical prediction $A_0\simeq 2.51$.}
\label{velocity}
\end{figure}

A snapshot of a typical configuration for a small system
with $k^*_H<k_0<k^*_\perp$  at
large times $\tau \gg  \tau_{cr}$  is shown in Fig.~\ref{config}.
  A shear flow with variation of the $u_x-$component in the
$y-$direction is observed.
Individual $v_x-$components of the particle velocities  are plotted in
Fig.~\ref{velocity} versus their $y-$coordinate. A fit to a sinusoidal curve
(solid line) shows that the solution described in (\ref{us}) is realized.

\begin{figure}
\begin{center}
      \epsfig{file=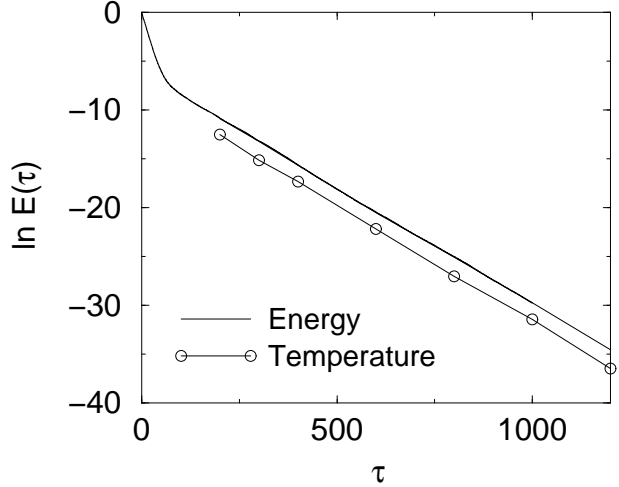,angle=270,width=8cm}
\end{center}
\caption{\small Energy per particle and temperature
as a function of number of collisions per particle
$\tau$. At $\tau=600$ one finds
$(E/\bar{T})_{\mbox{sim}}\simeq 5.38$
with a theoretical value $\gamma_0/{\cal D}_{\perp} k_0^2\simeq 4.14$}
\label{energyandtemperature}
\end{figure}

We assume that the number of collisions $\tau$  in the simulations
of Figs.~3 and 4 is sufficiently large, so that the components
$\tilde{\bf u}_{\bf k}$ are very close to their fixed point values.
Then, the relation, ${\cal D}_\perp
\overline{|\nabla\tilde{\bf u}|^2} =\frac{1}{2}d( \gamma_0 -{\cal D}_\perp
k_0^2)$, follows from (\ref{velstationary}) 
and (\ref{reduced}), 
and $\partial_{\tau}\ln\bar{T}\simeq-2{\cal D}_{\perp}k_0^2$.
The global temperature for large times, i.e. 
$\tau \gg \tau_{cr}=L^2/{\cal D}_{\perp}$, becomes,
\begin{equation}
\bar{T} \simeq T_e \exp ( -2 {\cal D}_\perp k_0^2 \tau ).
\label{tbar}
\end{equation}
Equation (\ref{tbar}) gives the temperature as a function of $\tau$. 
The integration constant {\it cannot} be determined from the theory,
but a fit of (\ref{tbar}) to the simulation data in Fig.~5 for
$\tau>200$ yields $T_e \simeq 3.87\times 10^{-4}$. We also note
that the mode coupling theory of Ref.~\cite{EPL-brito} gives the
same decay rate for the total energy, when this theory is
applied to a {\it small} system. In that case the Fourier sum
$(1/V) \sum_{\bf q}$  can not be replaced by $\int d {\bf
q}/(2\pi)^d$. On the contrary, the sum is essentially given by its first term
only.

The relation between $\tau$ and $t$ is defined through $ d\tau
=\omega(\bar{T})\,d t$, where the collision frequency $\omega$ is
proportional to the square root of $\bar T$, i.e.
\begin{equation}
\frac{d \tau}{d t}=\omega(\bar{T})
=\omega_0\sqrt{\frac{\bar{T}}{T_0}}
\simeq\omega_0 \sqrt{\frac{T_e}{T_0}}
\exp\left[-{\cal D}_{\perp}k_0^2\tau \right],
\label{omegabar}
\end{equation}
where $\omega_0=\omega(T_0)$ is the collision frequency
at the initial time.
As $\omega$ can be measured directly as a function of $t$
in MD simulations, it yields an independent determination of $T_e$
with the result $T_e\simeq 3.68 \times 10^{-4}$.
Integration of (\ref{omegabar}) yields,
\begin{equation}
\exp({\cal D}_{\perp}k_0^2 \tau)\simeq
{\omega_0} \sqrt{\frac{T_e}{T_0}}
     {\cal D}_{\perp}k_0^2 (t-t_e),
\label{time}
\end{equation}
valid for $(t-t_e)$ large. After eliminating
$\tau$ from Eq.~(\ref{tbar}) and
(\ref{time}) we obtain the global temperature  for
asymptotically large time $t$ as,
\begin{equation}
\bar{T} \simeq \frac{T_0}{(\omega_0 {\cal D}_{\perp}k_0^2)^2} 
\frac{1}{(t-t_e)^2},
 \label{realtimetemp}
\end{equation}
where visual inspection of Fig.~6 shows that $t_e\simeq 10^4$. 
The temperature shows {\em algebraic decay} 
with the same exponent as in Haff's law
(\ref{haffslaw}), but the prefactor in Haff's law does neither depend on the 
system size, nor on the dimensionality.
  In (\ref{realtimetemp}) it is proportional to $L^4 \sim N^{4/d}$,
whereas the prefactor in Haff's law is independent of the system
size (see Fig.6).

\begin{figure}[h]
\begin{center}
      \epsfig{file=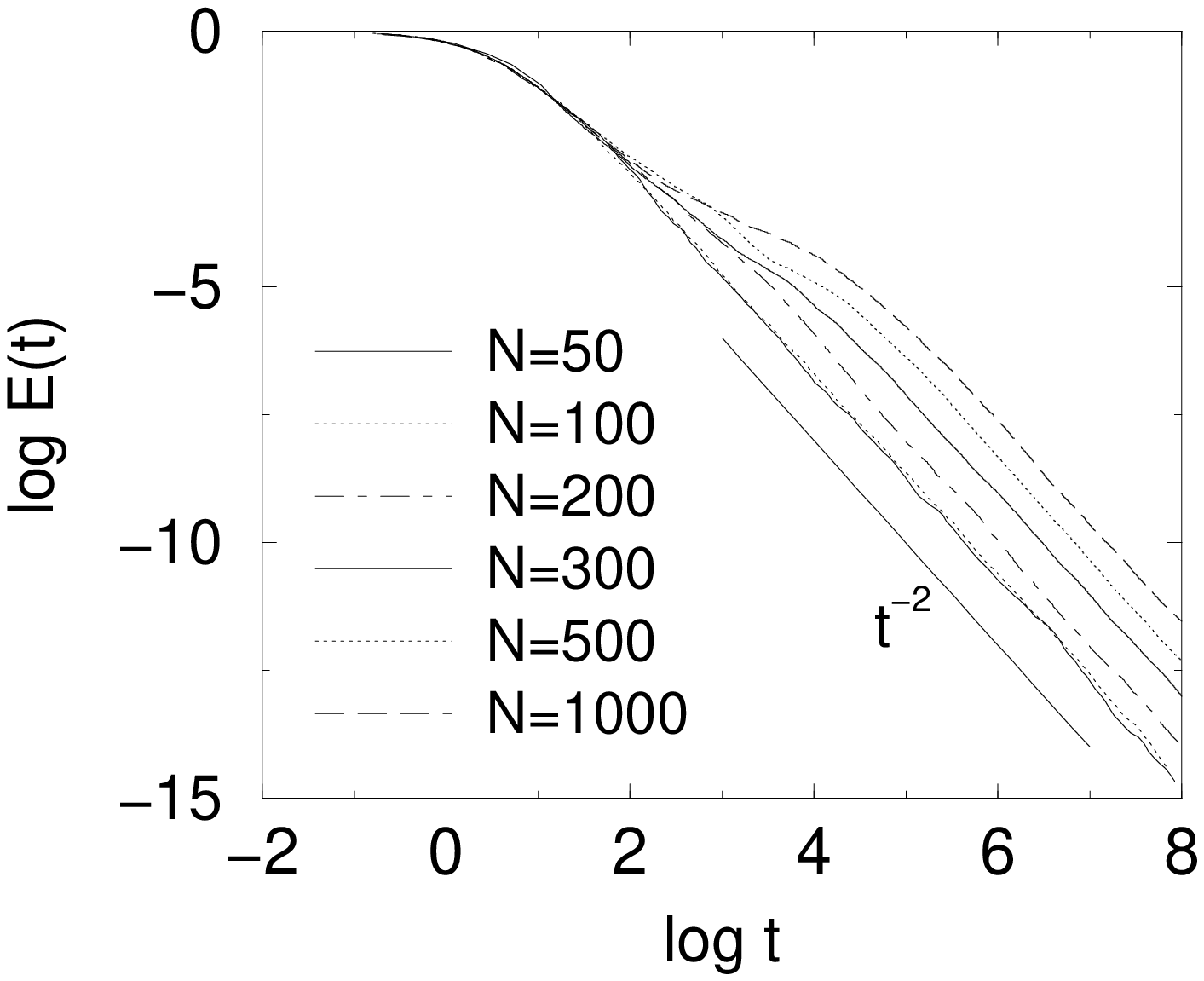,angle=270,width=6.3cm} \ \
      \epsfig{file=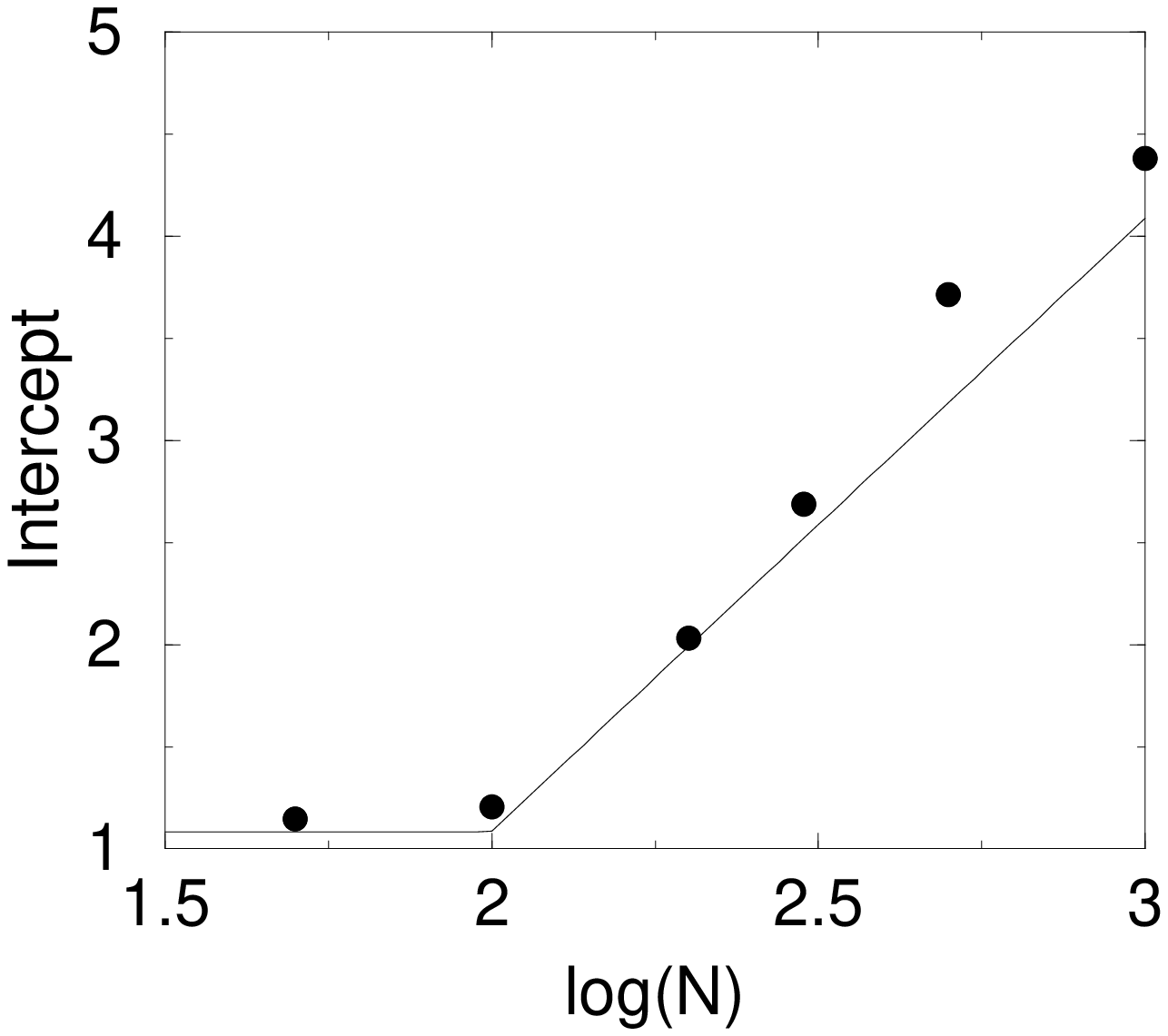,angle=270,width=5.7cm}
\end{center}
\caption{\small Left: Simulation results of the behavior
of $E(t)$ as a function of
$t$ for $N$ particles. At large times ($t \protect\mayapr 10^4$ for
$\phi=0.4$ and $\alpha=0.85$),
a power law tail $\sim 1/t^2$ is observed. The coefficient $B$,
defined in (\protect{\ref{energyt}}), is on average $2.3\times 10^{-5}$
with a theoretical value $B\simeq 1.2\times 10^{-5}$.
Right: the amplitude of the tail is proportional to $N^3$.
Here the solid line is the result of the theory,
Eq.~(\protect{\ref{energyt}}), while the dots are simulation
results extracted from the left panel.}
\label{temperatureint}
\end{figure}

Once we have $\bar{T}$, we can calculate the energy per particle,
defined in \cite{EPL-brito} as,
\begin{equation}
E=\frac{1}{n}\left[\,\overline{\frac{m}{2}n\,{\bf u}^2+
\frac{d}{2}n T} \,\right] =\bar{T}\,\left[\overline{\tilde{\bf u}^2}+
\frac{d}{2}\right],
\label{etot}
\end{equation}
where local density variations are supposedly small and
the relation ${\bf u} = \bar{v}_0 {\bf \tilde{u}}$
has been used. 
Combination of this result with (\ref{velstationary}) yields,
\begin{equation}
E \simeq  \frac{d\gamma_0}{2{\cal D}_{\perp}k_0^2}\,
\bar{T}.
\label{eandt}
\end{equation}
This shows that at long times,
the energy and temperature are proportional. Therefore,
energy decays exponentially when expressed in terms of
$\tau$ (see (\ref{tbar})), with the same decay rate
as the global temperature.
This property is observed by MD simulations, as shown in
Fig.~\ref{energyandtemperature}. From (\ref{eandt}) and
(\ref{realtimetemp})
the decay of the energy
per particle in terms of $t$
in the long time limit is given by
\begin{equation}
E(t)\simeq
 \frac{d\gamma_0\,T_0}{2\omega_0^2({\cal D}_{\perp}k_0^2)^3}\,\frac{1}{(t -
t_e)^{2}}
\equiv B \,\frac{N^{6/d}}{(t - t_e)^{2}}
.
\label{energyt}
\end{equation}

Hence, if we vary the number of particles $N$, at fixed packing
fraction $\phi$, the amplitude of $t^{-2}$-decay of the energy is
proportional to $N^{3}$ in two-dimensional systems, or to $N^{6/d}$
in $d$-dimensional systems. Comparison of Eq.~(\ref{energyt}) with
simulations is shown in Fig.~\ref{temperatureint}. It would also be
interesting to compare the coefficient $B$ in (\ref{energyt}) with
the simulation results of Ref.\cite{trizac-barrat} for the IHS
fluid which have been performed in five and six dimensions.

\begin{figure}[h]
\begin{center}
      \epsfig{file=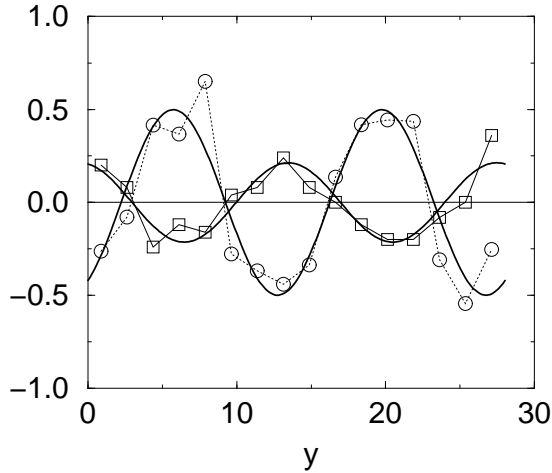,angle=270,width=8cm}
\end{center}
\caption{\small Density (squares) and temperature (circles)
 inhomogeneities,
$\delta \tilde{n}$ and $\delta \tilde{T}$ at $\tau=600$.
The solid lines correspond to nonlinear fits
to sinusoidal functions of half the period in
\protect{(\ref{localtemperaturedensity})}.
The simulated ratio of $\delta \tilde{T}$- to $\delta \tilde{n}$
amplitudes is here $2.0$, and the theoretical prediction is 2.3.}
\label{inhomogdensitytemperature}
\end{figure}

Finally, a systematic expansion to be presented in~\cite{wakou},
shows that the stationary solution of
local temperature and local density are
given by
\begin{eqnarray}
\delta\tilde{T}\equiv\frac{T(y)}{\bar{T}}-1
&\simeq & 
-\frac{{b}_T
V_0^2}{2d{\omega}} \cos[2(k_0 y +\theta_y)] 
\nonumber\\
\delta\tilde{n}\equiv\frac{n(y)}{\bar{n}}-1
&\simeq & 
\frac{
{b}_TV_0^2\,\bar{T}\left(\frac{\partial p}{\partial
T}\right)
}{
2d \omega\,\bar{n}\left(\frac{\partial p}{\partial n}\right)
}
\cos[2(k_0 y +\theta_y)],
\label{localtemperaturedensity}
\end{eqnarray}
where $\bar{n}=N/V$, and $\theta_y$ is the same phase factor
as given in (\ref{us}). It means that the density
and temperature inhomogeneities show a period which is half the
period of the shear flow profile. The relation between the spatial
periods has already been observed in Ref.\cite{4}
in MD simulations of a fluid of two-dimensional hard disks, and in
Ref.~\cite{brey-mariajose} in Direct Monte Carlo simulations of the
Boltzmann equation for an IHS gas.
Figure \ref{inhomogdensitytemperature} shows the observed
density and temperature inhomogeneities.
A fit to a sinusoidal curve supports the temperature and
density profiles given by (\ref{localtemperaturedensity}).

As shown in Eq.(\ref{localtemperaturedensity}), the temperature and density 
inhomogeneities
 are in opposite phase, implying
that dense regions are cold, and the dilute regions hot.
The amplitudes are such that the overall pressure is constant, as we have
assumed in the course of the paper.

\section{Conclusions}

Under the restrictions imposed in our derivations, 
as discussed in Sec.3, we have
shown that the unstable dynamics and formation of vortex 
patterns in the flow field of a freely evolving fluid
of inelastic hard spheres (IHS) can be cast in the
form of a time-dependent Landau-Ginzburg-type model
for a non-conserved order parameter. 
In the two-dimensional case (but not for $d\ge 3$)
the growth of the vortex pattern in 
the IHS fluid is qualitatively similar to 
spinodal decomposition for a non-conserved 
scalar order parameter, referred to as model A in
the Hohenberg--Halperin classification~\cite{5}.
The analogy between unstable IHS fluid 
in two-dimensions and spinodal 
decomposition has already been pointed out in Ref.~\cite{11}.
A difference between our model and model A is
that the energy functional (\ref{landauf}) contains a quartic term
with summations over two independent wave numbers. This
implies a non-local interaction of the order parameter
${\sf S}=\nabla \tilde{\bf u}$. 
The non-local interaction is caused by the scaled field 
$\tilde{\bf u}={\bf u}/\bar{v}_0\sim {\bf u}/\sqrt{\bar{T}}$.
Because the global temperature $\bar{T}$ is determined 
by ${\sf S}$ in all space (see Eq.(\ref{Tbar})),
the evolution of a local order parameter ${\sf S}$ is 
affected by ${\sf S}$ at any other points in space through $\bar{T}$.

We have shown that nonlinear 
viscous heating gives rise to a quartic term in the energy
functional, which is responsible for saturation of 
unstable vorticity modes. 
Unstable vorticity modes initially grow 
as predicted by the linear stability analysis,
but eventually saturate because of nonlinear viscous heating.
The influence of nonlinear viscous heating 
on the formation of temperature inhomogeneities and clustering 
has been pointed out by Goldhirsch and Zanetti~\cite{4},
and investigated in more detail by Brey et al.~\cite{brey-mariajose},
by comparing the results of a hydrodynamic model 
with viscous heating, using direct Monte Carlo simulation
of the Boltzmann equation.  

The theoretical predictions of our zeroth order theory 
for the flow field and the decay of the energy
are in quantitative agreement with MD simulations 
of small systems as shown in Sec.5.
They support the intuitive arguments used in deriving 
the zeroth order description, presented in Sec.3.
Moreover, the results presented here can be obtained 
as the lowest order approximation in a multi-time scale
expansion~\cite{wakou}.
This theory makes no assumption of incompressibility.
There are some properties of those small systems 
that can not be described by the zeroth order theory.
Density and temperature {\it inhomogeneities}
are only found in the next order of approximation.

It is worth mentioning that results of our zeroth and first order 
approximations are consistent with the results 
of a nonlinear analysis by 
Soto et al.~\cite{21}
of systems which are close 
to the stability thresholds $k^{*}_{\perp}$.
If the smallest wave number of the system $k_0$ is slightly smaller
than $k^{*}_{\perp}$, vorticity modes with the wave number $k_0$
grow so slowly that the rest of hydrodynamic modes are enslaved by
these vorticity modes. Besides, the amplitudes of these vorticity modes
at long times are expected to saturate because of nonlinear effects 
and remain small. On the basis of this consideration, they obtained
amplitude equations for the vorticity modes with $k_0$.
The inhomogeneous density and temperature are given as functions
of the vorticity modes with $k_0$.
 
Finally, we emphasize that the periodic boundary conditions,
with which the simulations were carried out, obviously 
play a crucial role in the formation of a shear flow profile
and of the density and temperature profiles,
presented in (\ref{us}) and (\ref{localtemperaturedensity}) respectively. 
Indeed, once the typical size of the
vortex patterns becomes comparable to the length $L$ of the system
for $\tau>\tau_{cr}=L^2/{\cal D}_{\perp}$,
the artificial periodic boundary conditions start to affect 
the evolution of the system.
Consequently, the long time regime $\tau\gg\tau_{cr}$
described by the stationary
solution is of less physical interest
than the regime of unstable growth $\tau\ll \tau_{cr}$,
where the typical size of the vortex patterns remains small
compared to the length $L$ of the system. 
A thermodynamically large system is always in the unstable growth regime.
Unfortunately, the only analytic 'large' time results for 
the unstable growth regime of granular fluids in 
{\it thermodynamically large} systems,
have been obtained from fluctuating (linear) hydrodynamic
equations or from mode coupling theory~\cite{EPL-brito,chen-physlett}, 
but not from truly nonlinear theories. An exception is a dilute gas of 
inelastic point particles in one dimension \cite{ben-naim}, 
where strong evidence supports the conjecture that 
the large space-time behavior is described by the adhesion model
and the Burgers equation.

\section{Acknowledgements}
M.E. acknowledges stimulating discussion with R. Desai and R. Kapral.
J.W. and R.B. acknowledge support of the foundation "Fundamenteel
Onderzoek der Materie (FOM)", which is financially supported by the Dutch
National Science Foundation (NWO). J.W. also acknowledges support of 
a Huygens scholarship. R.B. wants to thank the
Institute for Theoretical Physics of Universiteit Utrecht
for its hospitality. R.B. is supported
by grant DGES-PB97-0076 (Spain).

\end{document}